\documentclass[aps,prl,10pt,twocolumn,floatfix,superscriptaddress,preprintnumbers,shownopacs,showkeys,nofootinbib,notoccite,notitlepage]{revtex4-1}

\usepackage{graphicx}
\usepackage{dcolumn}
\usepackage{footmisc}

\usepackage{bm}
\usepackage{color}
\usepackage[dvipsnames]{xcolor}
\usepackage{amsmath}
\usepackage{float}
\usepackage[dvipsnames]{xcolor}
\definecolor{PRLblue}{RGB}{0,0,150}
\usepackage[
  colorlinks=true,
  linkcolor=PRLblue,
  citecolor=PRLblue,
  urlcolor=PRLblue
]{hyperref}
\begin{document}
\setlength{\abovedisplayskip}{5pt}
\setlength{\belowdisplayskip}{5pt}
\setlength{\abovedisplayshortskip}{5pt}
\setlength{\belowdisplayshortskip}{5pt}
\newcommand{\BD}{\textcolor{orange}}
\newcommand{\jtc}[1]{\textcolor{violet}{{\bf JTC}: [#1]}}
\newcommand{\mm}[1]{\textcolor{magenta}{{\bf MM}: [#1]}}
\newcommand{\DM}{{\rm DM}}
 \newcommand{\mchi}{m_{\chi}}
\newcommand{\figref}[1]{~Fig.~\ref{#1}}

\preprint{\texttt{FERMILAB-PUB-26-0453-T}}

\title{Galactic Center Neutrinos from Cosmic Ray-Dark Matter Interactions}

\author{Jorge Terol Calvo}
\email{terolcal@to.infn.it}
\affiliation{Istituto Nazionale di Fisica Nucleare, Sezione di Torino, I-10125, Torino, Italy}

\author{Pedro de la Torre Luque}
\email{pedro.delatorre@uam.es}
\affiliation{Departamento de F\'{i}sica Te\'{o}rica, M-15, Universidad Aut\'{o}noma de Madrid, E-28049 Madrid, Spain}
\affiliation{Instituto de F\'{i}sica Te\'{o}rica UAM-CSIC, Universidad Aut\'{o}noma de Madrid, C/ Nicol\'{a}s Cabrera, 13-15, 28049 Madrid, Spain}

\author{Mainak Mukhopadhyay}
\email{mainak@fnal.gov}
\affiliation{Astrophysics Theory Department, Theory Division, Fermi National Accelerator Laboratory, Batavia, Illinois 60510, USA}
\affiliation{Kavli Institute for Cosmological Physics, University of Chicago, Chicago, Illinois 60637, USA}

\author{Chris Cappiello}
\email{cappiello@wustl.edu}
\affiliation{Department of Physics and McDonnell Center for the Space Sciences, Washington University, St. Louis, MO 63130, USA}

\author{Gonzalo Herrera}
\email{gonzaloh@mit.edu}
\affiliation{Kavli Institute for Astrophysics and Space Research, Massachusetts Institute of Technology, Cambridge, MA 02139, USA}
\affiliation{Laboratory for Particle Physics and Cosmology, Harvard University, Cambridge, MA 02138, USA}


\begin{abstract}
The IceCube and ANTARES collaborations have recently reported evidence for high-energy neutrinos associated with the Galactic plane and the Galactic Ridge, offering a new pathway to search for dark matter (DM). 
Deep inelastic scattering of cosmic rays with sub-GeV DM in the Galactic halo produces a distinctive neutrino signature from meson decays.
Using detailed Galactic cosmic-ray maps and ANTARES observations, we derive 99\% C.L. upper limits on the DM–nucleon cross section that extend down to keV-scale masses. These results establish Galactic neutrino telescopes as a powerful, complementary probe of light DM, with substantial improvements expected from upcoming IceCube-Gen2 and KM3NeT observations.
\end{abstract}

    \maketitle

\section{Introduction}

The nature of dark matter (DM) remains one of the most compelling open questions in modern physics~\cite{Cirelli:2024}. Although the traditional paradigm has long focused on weakly interacting massive particles in the GeV–TeV mass range, increasing attention has shifted toward lighter candidates, particularly in the sub-GeV regime~\cite{Knapen:2017xzo,Essig:2022dfa,Mitridate:2022tnv}. This transition is motivated both by the absence of conclusive signals in conventional searches, mostly focused on direct detection experiments~\cite{Akerib:2022ort,Arcadi:2024ukq}, and by the theoretical plausibility of light DM scenarios~\cite{Lin:2011gj,Hochberg:2014dra,Hochberg:2014kqa,Essig:2015cda,Kuflik:2015isi,Knapen:2017xzo,CarrilloGonzalez:2021lxm,Cline:2024wja}. However, probing this low-mass frontier presents significant experimental challenges, as the energy deposited in terrestrial detectors becomes vanishingly small, rendering standard direct detection techniques largely insensitive.

In recent years, cosmic-ray (CR) interactions with light DM particles have been explored as a way to overcome the limitations of direct detection experiments. These interactions can induce a boosted DM component falling in the region of interest of such experiments 
 , see \textit{e.g.}~\cite{Bringmann:2018cvk, Wang:2021jic,Cappiello:2024acu, Gustafson:2025dff}, an anomalous CR cooling \cite{Cappiello:2018hsu,Herrera:2023nww,Gustafson:2024aom}, a detectable secondary flux of gamma-rays \cite{Hooper:2018bfw,Guo:2020oum, Ambrosone:2022mvk}, radio waves \cite{Hussein:2025llu}, and neutrinos \cite{Guo:2020oum, DeMarchi:2024riu, DeMarchi:2025xag}. The non-observation of either CR-DM boosting or cooling signatures, in particular from Active Galactic Nuclei, places leading constraints in the parameter space of MeV-scale DM, probing thermal freeze-out in some selected models.
Traditional DM models are usually dismissed phenomenologically for masses below a few MeV, as they would remain in thermal equilibrium during BBN, altering the neutron to proton yields \cite{Serpico:2004nm}. However, keV-scale DM may freeze in \cite{Hall:2009bx, Chang:2019xva} from an initial small abundance, and the corresponding combinations of couplings and masses that make up the observed relic abundance remain largely untested.

\begin{figure}[t!] 
    \centering 
    \includegraphics[width=0.49\textwidth]{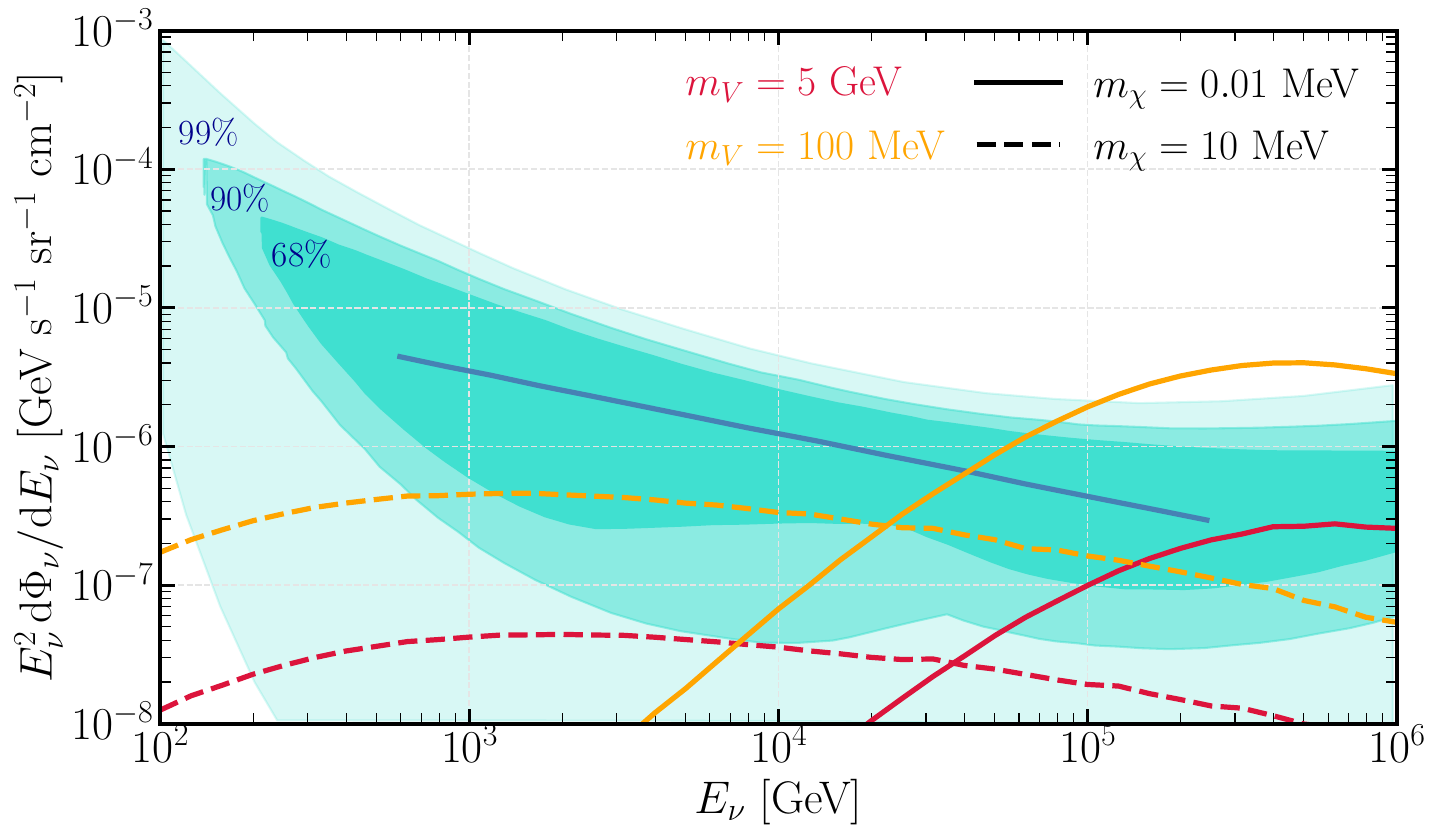}
    \caption{Expected neutrino signal coming from CR-DM inelastic interactions in the Galactic Ridge. We show the flux for a $m_{\rm DM}=10$ MeV (dashed) and a $m_{\rm DM}=10$ keV (solid) Dirac DM particle with a vector mediator with $m_V=5$ GeV (red) and $m_V=100$ MeV (orange), assuming couplings $G_D^2=g_q^2\cdot g_\chi^2=5$. The solid blue line corresponds to the ANTARES Galactic Ridge best-fit and the turquoise regions correspond to the 68\%, 90\% and 99\% C.L. contours~\cite{ANTARES:2022izu}.}

\label{fig:Spectra} 
\end{figure}

Recent observations have revealed the presence of a neutrino flux from the Galactic Plane, with an intensity roughly an order of magnitude lower than the total astrophysical neutrino flux. The detection and characterization of this diffuse neutrino emission associated with our Galaxy~\cite{ANTARES:2022izu,IceCube:2023ame} have opened a new observational window into the high-energy processes occurring in the Milky Way. This progress enables novel probes of physics beyond the Standard Model, including the possibility of detecting or constraining light DM, as we illustrate in Fig.~\ref{fig:Spectra}.

In this Letter we propose a promising way to test these sub-GeV DM models, through the neutrino yield arising from their deep-inelastic scatterings with CRs at the Galactic Center. Current measurements of the neutrino flux from the Galactic Ridge at ANTARES and future IceCube observations allow us to probe new parameter space of sub-GeV DM. The results presented here highlight the unique role of Galactic neutrino observations as a probe of light DM, opening a new window into previously unexplored regions of parameter space.

\section{ANTARES Galactic neutrinos}

The Galactic Ridge ($|l| < 30^\circ$, $|b| < 2^\circ$, in galactic coordinates) is a key region for studying high-energy neutrino emission in the Milky Way~\cite{ANTARES:2022izu, ANTARES:2025wvi, DeLaTorreLuque:2025zsv}. In this region, neutrinos are expected to arise mainly from hadronic interactions of CRs — primarily protons and helium — with interstellar gas. ANTARES provides the first robust constraints on neutrino emission from this region using 13 years (2007–2020) of data. To extract the signal, the collaboration used power-law templates fitted across six independent energy bins~\cite{ANTARES:2022izu, ANTARES:2025wvi}. Systematic uncertainties associated with the assumed spectral index were evaluated by varying $\alpha$ between 1 and 4, producing large error bands. This approach yields a largely data-driven characterization of the Ridge emission with minimal model dependence. Therefore, such robust upper limits in the neutrino flux at the Ridge allows us to set limits on the neutrino production from inelastic scattering of DM with CRs.

Interestingly, ANTARES measurements indicate a relatively hard neutrino spectrum in the inner Galaxy compared to that inferred from local CRs, consistent with earlier hints from $\gamma$-ray observations of the Ridge~\cite{TIBET, ANTARES:2022izu}. The agreement with $\gamma$-ray–calibrated diffuse emission models supports a CR-interaction-dominated origin and helps constrain uncertainties in the CR distribution in the inner Galaxy~\cite{ANTARES:2025wvi, DeLaTorreLuque:2025zsv}. In particular, this may reflect spatially varying CR transport properties~\cite{TIBET, Gaggero2015, Cerri_2017, DelaTorreLuque:2022ats}, but it might be related also to unexpected contributions in the inner Galaxy, including new physics~\cite{Zuriaga-Puig:2026mfz}. Given the higher DM and CR densities in the inner regions, the Ridge is also a natural target for indirect DM searches, where the rate of CR–DM interactions is expected to be particularly high, potentially contributing to the observed neutrino flux through deep inelastic scattering.

\section{Galactic cosmic-ray nuclei distribution}

Recent CR experiments have provided highly precise measurements of CR nuclei spectra in the GeV–TeV range. Although these observations are limited to local fluxes, a combination of gamma-ray and CR data allows us to improve the description of their distribution across the Galaxy~\cite{DeLaTorreLuque:2025zsv}. In this work, we consider a range of scenarios to capture uncertainties in the CR distribution, enabling us to quantify their impact on our results. On one end, we adopt a conservative setup — referred to as the Minimal (or Base–Minimal) scenario — based on fits to local CR data under the assumption of uniform diffusion throughout the Galaxy (See Ref.~\cite{Luque:2022buq} for more details). On the other end, we explore the so-called $\gamma$-optimized transport scenarios~\cite{DeLaTorreLuque:2025zsv}, which provide a physically motivated framework to reconcile $\gamma$-ray and Galactic neutrino observations across a wide energy range with local CR measurements~\cite{Luque:2022buq, DelaTorreLuque:2022ats, DeLaTorreLuque:2025zsv, Tovar-Pardo:2024cbp}. This scenario reproduces observations by Fermi-LAT of a progressive hardening of the diffuse $\gamma$-ray emission towards the center of the Galaxy~\cite{Gaggero:2014xla, Gaggero:2015xza, Fermi-LAT:2016zaq, Yang2016prd, Lipari:2018gzn, Pothast:2018bvh}, adopting a non-uniform diffusion coefficient. Compared to conventional (spatially uniform) diffusion scenarios, $\gamma$-optimized models show a clear improvement in the inner Galaxy, where they capture both the normalization and the spectral shape of the observed emission.

Within this framework, Ref.~\cite{Luque:2022buq} derived two variants that allow us to bracket the uncertainties associated with very high-energy CR observations (see Fig.~10 of Ref.~\cite{DeLaTorreLuque:2025zsv}). These configurations were denoted as the $\gamma$-optimized Min and Max setups. As shown in Ref.~\cite{DeLaTorreLuque:2025zsv}, the $\gamma$-optimized Min setup is particularly compelling: by adopting a softer CR injection spectrum above $\sim$10 TeV, motivated by features observed in DAMPE and CREAM data, it provides a conservative yet robust description of the high-energy CR population. Importantly, this setup not only remains compatible with $\gamma$-ray data, but also aligns well with current neutrino constraints and hints from experiments such as IceCube~\cite{IceCube:2023ame}. In fact, the predicted neutrino flux associated with the $\gamma$-optimized Min model is consistent with the level of the diffuse Galactic component suggested by IceCube~\cite{DeLaTorreLuque:2025zsv}, making it a natural benchmark for multi-messenger studies. 

\section{Neutrino flux from cosmic ray-dark matter interactions}

In the scenario that we are testing, high-energy neutrinos arise from interactions between CRs and DM particles in the Galactic halo through deep inelastic scattering. Here, the resulting neutrino flux is computed by performing a line-of-sight integral over the region of interest (i.e. the Ridge), obtaining a differential flux given by 
\begin{align}
    \frac{d\phi_{\nu}}{d\Omega dE_{ \nu}} (l, b, E)=  &\int ds \,\frac{\rho_{\DM}}{m_{\chi}}  \int dE_{\rm CR}  \,\frac{d\phi_{{\rm CR}}(E_{\rm CR})}{d\Omega \, dE_{\rm CR}} \notag\\& \,\sigma_{\rm CR\chi}(E_{\rm CR})\, \frac{dN_\nu}{dE_{\nu}}\Big(E_{\rm CR}\Big)\, ,
    \label{eq:TotalInt}
\end{align}
where $l$ and $b$ are the Galactic longitude and latitude, respectively, and $s$ is the distance along the line of sight. In our fiducial case, the DM density profile $\rho_\DM$ is modeled following a standard Navarro-Frenk-White (NFW) profile~\cite{Navarro:1995iw, Navarro:1996gj, Navarro:1997}, with a scale radius $R_s = 20$~kpc~\cite{Cirelli:2011, Cirelli:2024} and a local density of $\rho_\odot = 0.4 \ \mathrm{GeV} \mathrm{cm^{-3}}$ \cite{Benito_2019, 2021PDU....3200826B}. Notably, our predicted limits weaken by just a factor of 2–3 even when adopting a highly conservative DM distribution, such as a Burkert profile~\cite{Burkert:1995yz} (see Supp. Material). We sum the contributions from CR H and He, whose spatial and kinematic distribution, $d\phi_{\rm CR}/dE_{\rm CR}$, is described by the $\gamma$-optimized Min model. Full details on the impact of astrophysical uncertainties including the CR distribution and the DM profile are provided in the Supp. Material, where we find that they introduce an uncertainty of a factor of a few. We remark that this is a substantial improvement with respect to previous studies that adopt a spatially constant CR distribution based on spectral parametrizations of local data (commonly only based on AMS-02 data; while the models used here use a combination of all available local CR and gamma-ray data). This is key to exploiting the full potential of Galactic Center neutrino observations

The cross section $\sigma_{\rm CR\chi}$ and the rate $dN_\nu/dE_{\nu}$ govern the energy spectrum of secondary neutrinos originating from meson decays after the inelastic scattering of a CR nucleus off a DM particle, $\chi$. Here, we define the cross sections as $$\sigma_{\rm CR\chi} = A^{2/3}\cdot G_D^2 \cdot\hat\sigma(E_\mathrm{CR}, m_\chi, m_{\rm med})$$ where $A^{2/3}$ is the usual mass-number scaling for nuclei, $G_D^2 = g_p^2 \, g_{\chi}^2$ is a dimensionless, energy-independent combination of the couplings and $\hat\sigma(E_\mathrm{CR}, m_\chi, m_{\rm med})$ encodes the full kinematic dependence on the center-of-mass energy and mediator propagator. The neutrino energy spectrum per event will be given by $dN_\nu/dE_\nu$. To robustly predict this yield across disparate kinematic regimes, we utilize a Monte Carlo simulation pipeline: the underlying hard scattering events are generated using \textsc{MadGraph5\_aMC@NLO}~\cite{Alwall:2014hca}, while parton showering, hadronization, and subsequent particle decays are captured by \textsc{Pythia 8}~\cite{Bierlich:2022pfr}. This setup allows us to track the energy fractions transferred to resulting neutrinos realistically, even down to the sub-GeV DM mass regime. By convolving these differential particle yields with the realistic spatial CR and DM distribution in the inner Galaxy, we obtain the expected diffuse neutrino morphology and spectrum for the Galactic Ridge. Most of the previous studies with gamma rays as observable adopted simple constant cross sections convolved with p-p photon production. For comparison, we also show the case of constant cross sections in the Supp. Material.

The predicted neutrino flux from the Ridge from CR-DM inelastic interactions is compared with ANTARES observations in Fig.~\ref{fig:Spectra}, for the cases of a DM particle with mass of 10 keV (solid lines) and 10 MeV (dashed lines), in the two benchmark mediator masses adopted. This illustrates the potential of these measurements to probe a vast range of DM masses.

\section{Limits on light dark matter}

We adopt a simplified model comprising a Dirac DM particle, $\chi$, interacting with Standard Model quarks via a vector mediator, $V$, as a benchmark scenario. The model is described by the Lagrangian
\begin{equation}
    \mathcal{L} \supset 
    g_\chi\,\bar\chi\gamma^\mu\chi\, V_\mu 
    + g_q \sum_q \bar q\gamma^\mu q\, V_\mu 
    + \frac{1}{2}m_V^2\,V_\mu V^\mu \,,
    \label{eq:lagrangian}
\end{equation}
where $g_\chi$ and $g_q$ are the DM and (universal) quark couplings to the  mediator, respectively. 

To derive upper limits on the DM-nucleon scattering cross section, the predicted secondary neutrino flux from the Galactic Ridge is confronted with the recent measurements by the ANTARES collaboration~\cite{ANTARES:2022izu}. A conservative methodology is employed, requiring that the theoretical expected neutrino flux does not exceed the 99\% confidence level (C.L.) upper limits reported by ANTARES in any observational energy bin. Specifically, the expected differential neutrino flux at energy $E_i$ is restricted by the corresponding observed limit, $\Phi_{\text{obs}, i}^{99\%}$
\begin{equation}
    \int_{E_{\text{min}, i}}^{E_{\text{max}, i}} \frac{d\phi_{\nu}}{d\Omega dE_{\nu}} dE_{\nu} \leq \Phi_{\text{obs}, i}^{99\%} \, .
    \label{eq:limit_condition}
\end{equation}

\begin{figure}[t!] 
    \centering 
    \includegraphics[width=0.48\textwidth]{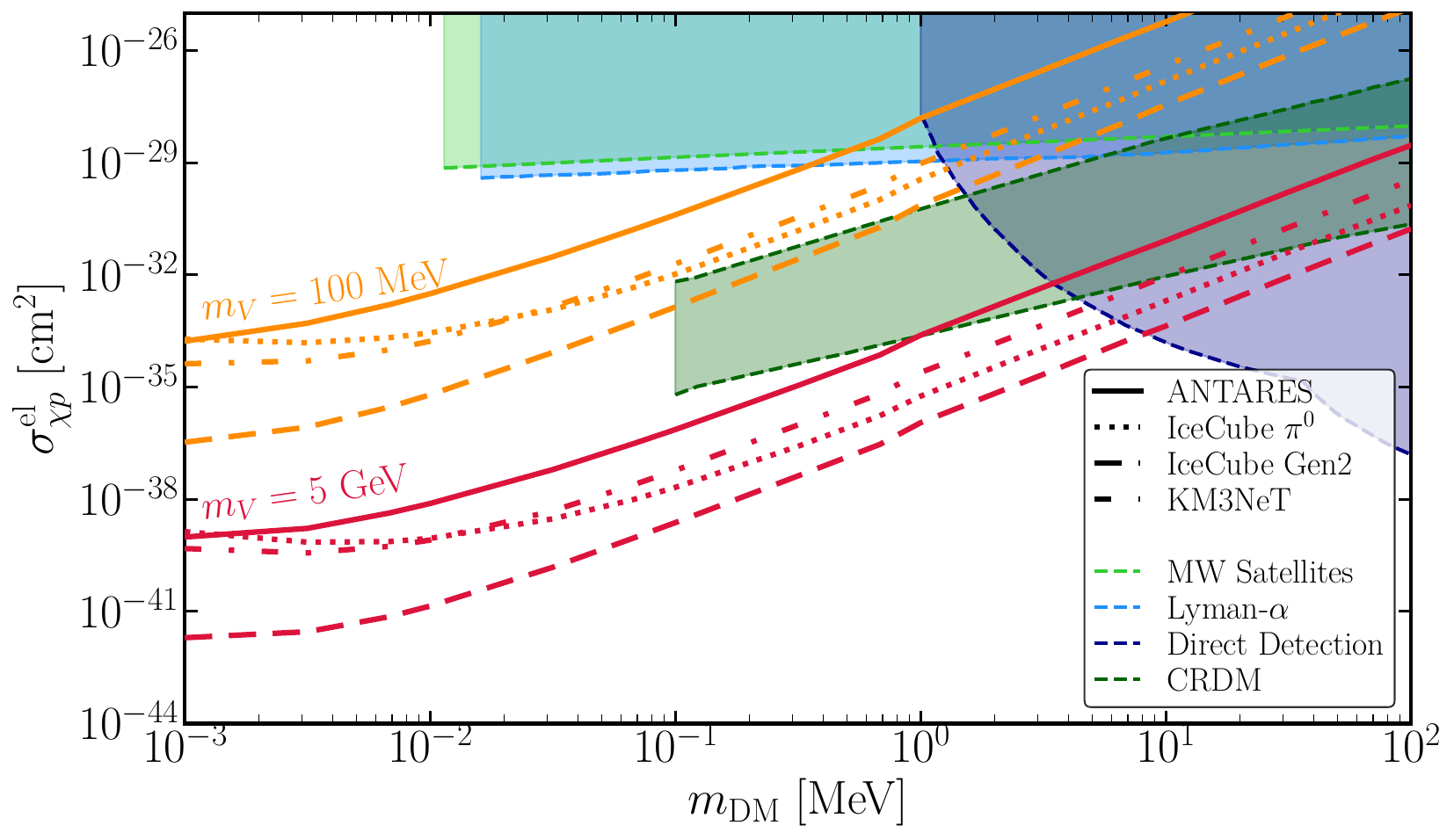}
    \caption{Derived $99\%$ C.L. upper limits on the DM-nucleon elastic scattering cross section as a function of the DM mass, utilizing the ANTARES neutrino observations in the Galactic Ridge and prospects derived from IceCube observations of the galactic plane, IceCube-Gen2 and KM3NeT. In addition we show limits coming from Lyman-$\alpha$ forest observations~\cite{Rogers:2021byl}, Milky Way satellite galaxy population~\cite{DES:2020fxi}. We also show limits coming from Direct Detection including PandaX-4T~\cite{PandaX:2023xgl}and DAMIC-M~\cite{DAMIC-M:2025luv} and finally the limits coming from CRDM obtained with XENON1T~\cite{Dent:2019krz}. Other constraints that might be of interest are discussed in the main text.}
\label{fig:Ridge_Consts} 
\end{figure}

The parameter space is systematically scanned spanning a wide range of DM masses, $\mchi$, for different choices of mediator masses. For each mass configuration, the fiducial interaction coupling $G_D^2\equiv (g_\chi\cdot g_q)^2$ is iteratively scaled until the condition in Eq.~\eqref{eq:limit_condition} is achieved. This procedure directly yields the 99\% C.L. upper bounds on the elastic DM-nucleon cross section, shown in \figref{fig:Ridge_Consts}, across the sub-GeV mass regime, using the relation~\cite{Guo:2020oum,Wang:2025ztb}
\begin{equation}
    \sigma^{\rm el}_{\chi p} 
    \equiv \frac{g_p^2\,g_\chi^2\,\mu_{\chi p}^2}{\pi\,m_V^4}
    = G_D^2\, \frac{9\,\mu_{\chi p}^2}{\pi\,m_V^4}\,,
    \label{eq:sigma_NR}
\end{equation}
where $\mu_{\chi p} = m_\chi m_p/(m_\chi + m_p)$ is the DM--proton reduced mass. Thus, it provides independent constraints derived entirely from the Galactic high-energy neutrino sky.

\subsection{Existing Constraints}

Comparison with other constraints is not straightforward as assumptions and benchmark models differ from one study to another. Here we review existing constraints and their validity against our benchmark model and show them accordingly in \figref{fig:Ridge_Consts}.

\textbf{Model-Independent -}~There exist several constraints on light DM that apply whether we assume a constant cross section, an arbitrarily heavy vector mediator, or a mediator comparable in mass to the DM. Constraints from the Lyman-$\alpha$ forest~\cite{Rogers:2021byl} and the Milky Way satellite galaxy population~\cite{DES:2020fxi} extend down to approximately 10 keV. While these constraints rely entirely on scattering, there are also very strong constraints on light DM from $N_{\rm eff}$, assuming that the DM is either produced thermally, or interacts strongly enough to thermalize after production; in e.g., Ref.~\cite{Krnjaic:2019dzc} cross sections are bounded down to $10^{-49}\ \mathrm{cm}^2$ for masses below a few MeV. In addition, fermionic dark matter is bounded to be heavier than $\sim 1\ \mathrm{keV}$ using the Pauli principle and observations of dwarf spheroidal galaxies~\cite{Alvey:2020xsk}. Strong gravitational lensing and the Milky Way satellite galaxy population similarly exclude warm dark matter lighter than around 10 keV~\cite{Nadler:2021dft}.

While direct detection bounds mostly lose sensitivity below DM masses around 100 MeV, constraints based on the Migdal effect in semiconductors can extend down to around 1 MeV~\cite{SENSEI:2020dpa, Berghaus:2022pbu, SENSEI:2023zdf, DAMIC-M:2025luv}. Liquid noble detectors set stronger limits for masses above about 30 MeV~\cite{XENON:2019zpr,DarkSide:2022dhx,PandaX:2023xgl} (see also Ref.~\cite{Xu:2026acq}).

\textbf{Heavy Vector Mediator -}~A common assumption is that of a simplified model involving a heavy mediator. We can compare our results with results from CR-boosted or blazar-boosted DM which assume a heavy vector mediator. For CR-boosted DM, Ref.~\cite{Dent:2019krz} reports limits from XENON1T, Ref.~\cite{LZ:2025iaw} from LZ and Ref.~\cite{COSINUS:2026acs} from COSINUS, while Ref.~\cite{Cappiello:2024acu} reports similar limits from IceCube. The scalar mediator case is slightly more widely considered~\cite{Bondarenko:2019vrb,Ambrosone:2022mvk,Super-Kamiokande:2022ncz,Cappiello:2024acu}; although limits on the scalar mediator case are not strictly applicable to the model we consider, similar bounds could likely be derived for the case of a vector mediator. Bounds for both vector and scalar mediators have also been derived from cosmic ray cooling in NGC 1068 \cite{Herrera:2023nww,Gustafson:2024aom, Mishra:2025juk, Chun:2026muh}.

\textbf{Constant Cross Section -}~Many constraints based on CR boosting of DM have been set by assuming a constant (i.e. energy-independent) total DM-proton cross section (e.g. Ref.~\cite{Bringmann:2018cvk}). The strongest of these come from Super-Kamiokande~\cite{Super-Kamiokande:2022ncz} and from LZ data~\cite{Maity:2022exk}. Stronger limits may be set based on DM interactions with AGNs/blazars~\cite{Wang:2021jic,Herrera:2023nww}. Additional bounds also come from the effect that this scattering would have on the observed CR spectra~\cite{Cappiello:2018hsu,Meighen-Berger:2025hrq}. Future gamma-ray observations can also constrain these CR-DM interactions~\cite{Reis:2024wfy}. We refer the reader to the supplemental material for a comparison of our bounds assuming a constant cross section and existing ones.

\textbf{Light Mediators -}~Separately from DM itself, there exist many constraints on new mediators that might couple DM to nucleons~\cite{Knapen:2017xzo,Bauer:2018onh,Bell:2023sdq}. While a full exploration of the possible models is beyond the scope of this work, we note that these constraints tend to become weaker for mediator masses above about 100 MeV. For many models, there also exists viable parameter space above both supernova constraints (due to these mediators and their decay products being unable to escape the supernova) and displaced vertex searches (due to the short decay length at large couplings). As we take 100 MeV and 5 GeV as our benchmark values for the mediator mass, stellar cooling and fifth force constraints are not applicable to these models, and in the heavier case even supernova cooling and beam dump experiments lose sensitivity.

\subsection{Potential of upcoming neutrino observations: IceCube, IceCube-Gen2 and KM3NeT}
In addition to the ANTARES analysis, we consider preliminary observations of the Galactic Plane ($|l| \leq 180^\circ$ and $|b| \leq 90^\circ$) by IceCube~\cite{IceCube:2023ame}. Unlike the ANTARES observations, the IceCube results use various templates of GeV gamma-rays to report a more significant but model-dependent diffuse neutrino flux. Therefore, as compared to the ANTARES constraints, the ones obtained from the IceCube are more stringent, but less robust as a result of being model-dependent. 
To illustrate the IceCube constraints we use the energy spectra of the best-fit neutrino flux between $1$ TeV to $70$ TeV for the $\pi^0$ model, and compare with our flux (see Supp. Material for more details).

We highlight the prospects for next-generation neutrino detectors such as KM3NeT and IceCube-Gen2. The former is a km$^3$ detector in the Mediterranean Sea consisting of two detectors ARCA and ORCA. IceCube-Gen2 on the other hand, is the successor of IceCube, with an effective area roughly $10^{2/3}$ times bigger than IceCube. We scale the current observations in a conservative way to obtain the respective constraints for KM3NeT~\cite{LeBreton:2019lpq} and IceCube-Gen2~\cite{IceCube-Gen2:2020qha}.

To derive the prospects for KM3NeT, we use the fact that the effective area of ARCA is roughly $10 - 100$ times greater than that of ANTARES~\cite{Bozza:2023vO}. This allows us to scale down the constraints by a factor of $\sim 3.5$ assuming the lower end of the scaling. For IceCube-Gen2, we extend the current IceCube best-fit neutrino flux across a wider energy range from $0.5$ TeV to $1000$ TeV and compute the $90$\% interval to then obtain the limits. Given that IceCube-Gen2 has a factor of $\sim 5$ increase in the effective area over IceCube, we rescale the limits accordingly. Note that for both detectors the realistic improvement for the detector sensitivities will be better owing to better energy and directional reconstruction techniques. Thus, our projected limits are conservative in this regard.

We show the prospects in Fig.~\ref{fig:Ridge_Consts}, apart from the overall improvement, we note that for IceCube-Gen2  we also note that the limits become better for $m_{\rm DM} \lesssim 0.01$ MeV. This is due to the inclusion of a wider energy range for the neutrino flux.


\section{Conclusions}

In this work, we have explored the production of neutrinos from the inelastic interaction of high-energy CRs with DM particles in the Milky Way. Such interactions provide a novel mechanism for neutrino production that can be probed by neutrino telescopes such as ANTARES and IceCube, particularly in the case of sub-GeV DM.

A key aspect of our analysis is the use, for the first time, of neutrino observations from the Galactic Center region to probe cosmic ray-DM interactions. These observations correspond to fluxes that are more than an order of magnitude lower than the total astrophysical neutrino flux, requiring an accurate modeling of the underlying CR distribution and associated astrophysical backgrounds. To address this challenge, we have implemented an improved description of the Galactic CR population and performed a comprehensive treatment of the astrophysical uncertainties affecting the predicted signal. Such a conservative treatment reduces potential background mismodeling and enhances the robustness of the search for exotic contributions to the neutrino flux, including those arising from DM interactions.

We find that Galactic Ridge observations allow us to probe DM masses in the range from approximately $\sim 1$~keV up to nearly $1$~GeV, covering a particularly interesting region of parameter space that remains largely unexplored by conventional direct detection experiments. The resulting constraints improve upon current direct detection limits, extend sensitivity to lower DM masses than those typically accessible through cosmological probes, and are competitive with, or stronger than, existing indirect constraints in some regions of parameter space.

Our analysis has focused on Dirac DM and vector mediators with masses above $100$~MeV, motivated by the strong constraints that generally apply to the production of light mediators in stellar media and in collider experiments \cite{Knapen:2017xzo, Bell:2023sdq}. Nevertheless, we expect our conclusions to remain largely unchanged for alternative DM candidates, such as scalar or Majorana particles, as the velocity suppression in the cross section is compensated by the large cosmic energies. Similarly, replacing the vector mediator by a scalar or pseudoscalar mediator is not expected to significantly modify the overall phenomenology or the resulting constraints.

Importantly, our constraints probe DM interactions in the unique environment of the Galactic Center and rely on entirely different systematic uncertainties and underlying assumptions than other existing probes. We have further shown that our bounds can be significantly improved in the future. A combined analysis of neutrino data, including a detailed treatment of the background contribution from CR interactions with the interstellar medium, can improve the present ANTARES limits by at least an order of magnitude. In addition, the same CR--DM inelastic interactions are expected to produce a correlated $\gamma$-ray flux, primarily from neutral pion decays. However, there are no published diffuse $\gamma$-ray observations from the Galactic Center in the TeV-PeV range. A joint multimessenger analysis combining Galactic neutrino and $\gamma$-ray observations could further strengthen these constraints and disentangle potential DM signals from astrophysical backgrounds; this will be explored in future work. Better angular resolution toward the Galactic Center in future neutrino telescopes will further reduce background levels and help isolate a potentially larger DM-induced neutrino flux, owing to the higher expected central dark matter density \cite{Gondolo:1999ef}.

Looking ahead, future observations from IceCube and next-generation neutrino observatories will substantially enhance the sensitivity to neutrinos produced in CR--DM interactions. Such improvements may lead to constraints that become competitive with, or even surpass, those obtained from collider experiments. More broadly, Galactic neutrino observations constitute a powerful and complementary probe of particle DM in one of the most informative regions of the Galaxy, opening a promising avenue for the exploration of light DM and other forms of new physics.

\section*{Acknowledgments}
\noindent
JTC wants to thank Andrea Giovanni De Marchi for helpful discussion regarding the PYTHIA implementation. JTC acknowledges support from the  Research grant TAsP (Theoretical Astroparticle Physics) funded by \textsc{infn}, from the Italian Ministry of University and Research (MIUR) via the PRIN 2022 Project No. 2022F2843 “Addressing systematic uncertainties in searches for dark matter” and from the European Research Area (ERA) via the UNDARK project (project number 101159929).
PDL has been supported by the Juan de la Cierva JDC2022-048916-I grant, funded by MCIU/AEI/10.13039/501100011033 European Union "NextGenerationEU"/PRTR, and is currently supported by Ramón y Cajal RYC2024-048445-I grant, which is funded by MCIU/AEI/10.13039/501100011033 and FSE+. The work of PDL is also supported by the grants PID2021-125331NB-I00 and CEX2020-001007-S, both funded by MCIN/AEI/10.13039/501100011033 and by ``ERDF A way of making Europe''. PDL also acknowledges the MultiDark Network, ref. RED2022-134411-T.
M.\,M acknowledges support from the FermiForward Discovery Group, LLC under Contract No. 89243024CSC000002 with the U.S. Department of Energy, Office of Science, Office of High Energy Physics.
CVC was generously supported by Washington University in St. Louis through the Edwin Thompson Jakeynes Postdoctoral Fellowship. 
The work of GH was supported by the Neutrino Theory Network Fellowship with contract number 726844.
We acknowledge the support of the European Consortium for Astroparticle Theory in the form of an Exchange Travel Grant. This article is based upon work from COST Action COSMIC WISPers CA21106, supported by COST (European Cooperation in Science and Technology).

\bibliography{references}
\clearpage
	\onecolumngrid
\begin{center}
  \textbf{\large Supplemental Material for Galactic Center Neutrinos from Cosmic Ray-Dark Matter Interactions}\\[.2cm]
  \vspace{0.05in}
  {
Jorge Terol Calvo, Pedro de la Torre Luque, Mainak Mukhopadhyay, Chris Cappiello, Gonzalo Herrera
}
\end{center}
\vspace{1 cm}
	\twocolumngrid

\section{Cross sections}

In the main text, we present bounds based on a benchmark model where the dark matter is a Dirac fermion interacting with Standard Model quarks via a vector mediator. The kinematic dependence of the deep inelastic scattering cross section is sensitive to the relationship between the dark matter mass, $m_\chi$, and the mediator mass, $m_V$. For our primary results, we investigate models with fixed mediator masses: $m_V = 5$ GeV and $m_V = 100$ MeV. Fixing the mediator mass modifies the momentum-transfer dependence of the propagator, which consequently alters the scaling of the cross section with the incident cosmic-ray energy. 

\begin{figure*}[h!] 
    \centering 
    \includegraphics[width=0.49\textwidth]{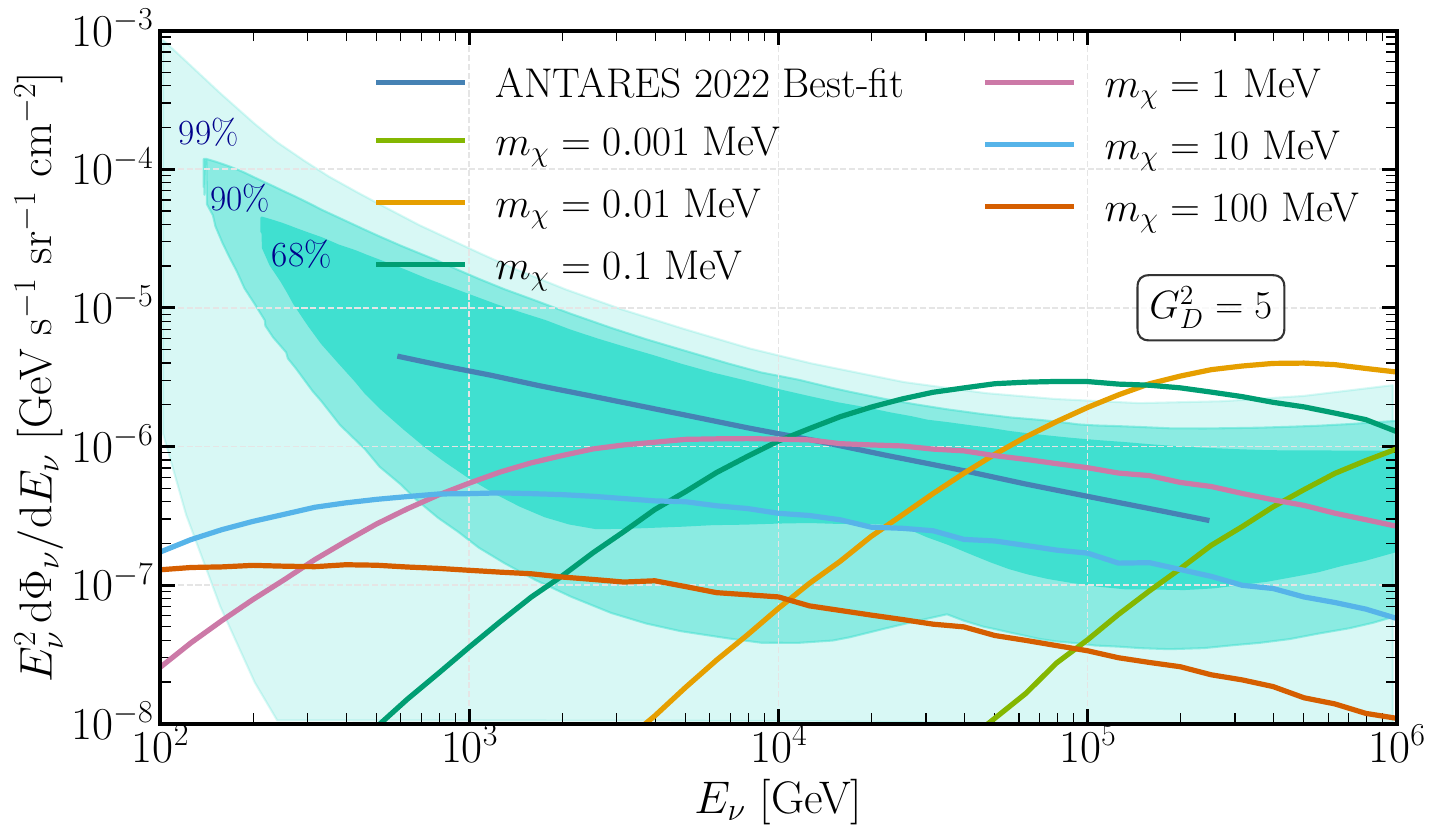}
    \includegraphics[width=0.49\textwidth]{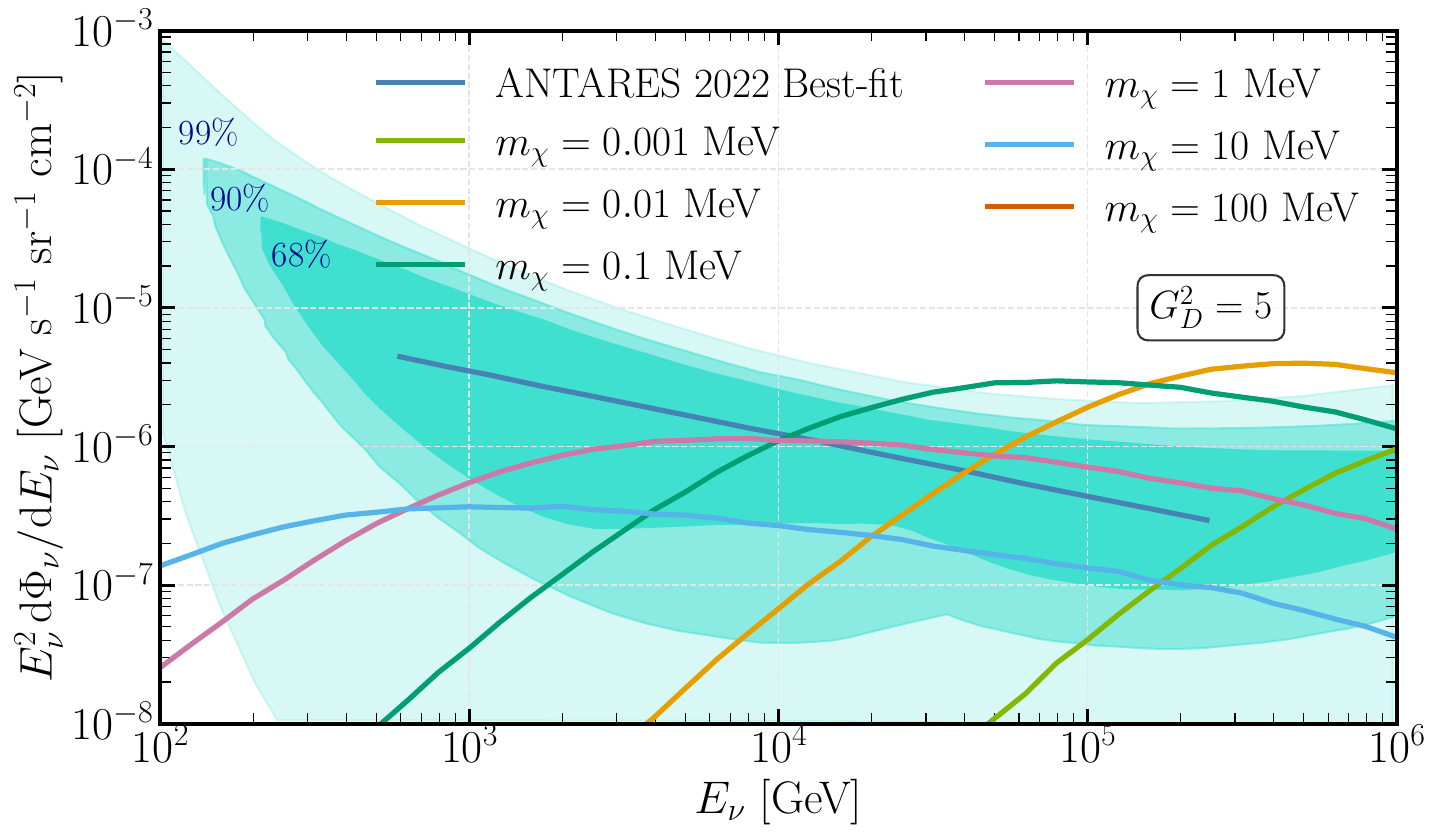}
    \includegraphics[width=0.49\textwidth]{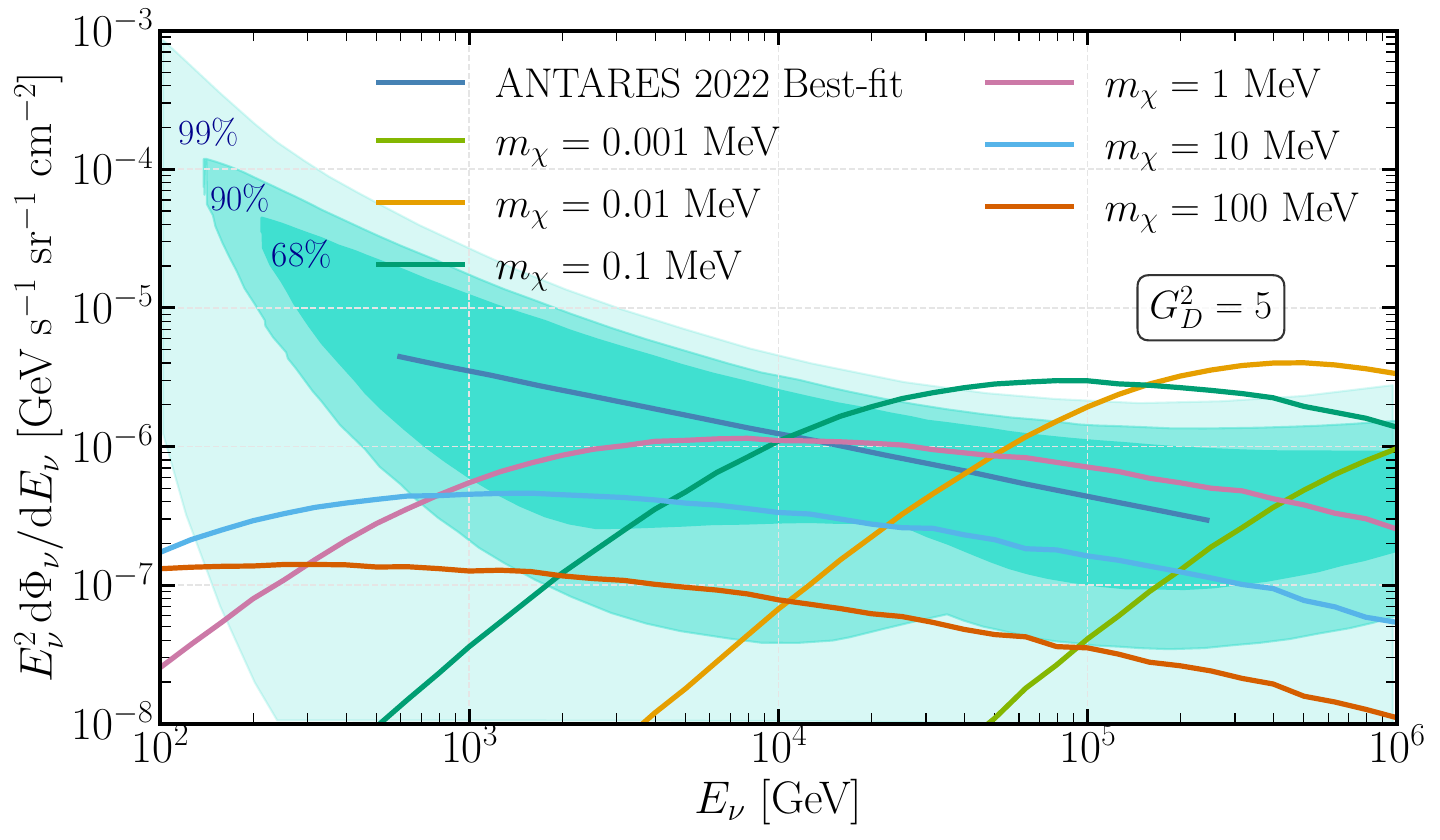}
    \includegraphics[width=0.49\textwidth]{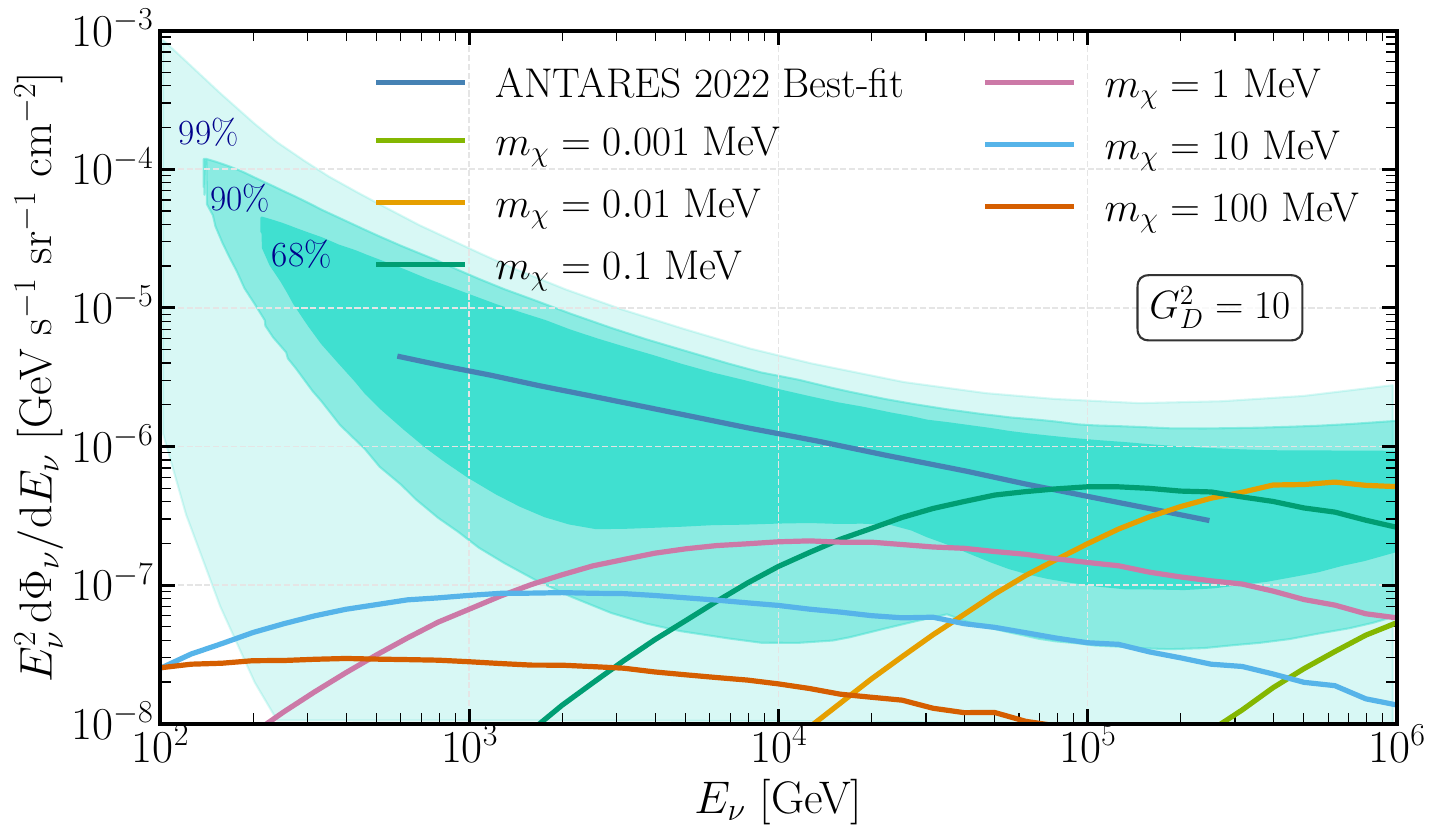}
    \caption{Expected neutrino fluxes from CR-DM interactions coming from the galactic Ridge under different DM mass-mediator mass relation. The value of the couplings squared $G_D^2$ is shown. \emph{Top Left: }$m_V = 3\, m_\chi$. \emph{Top Right: } $m_V = 100\, m_\chi$. \emph{Bottom Left: } Fixed mediator mass at m$_{V}= 100$ MeV. \emph{Bottom Right: } Fixed mediator mass at m$_{V}= 5$ GeV.}
\label{fig:Spectra_Cross_Section_Medmass} 
\end{figure*}

\begin{figure*}[h!] 
    \centering 
    \includegraphics[width=0.49\textwidth]{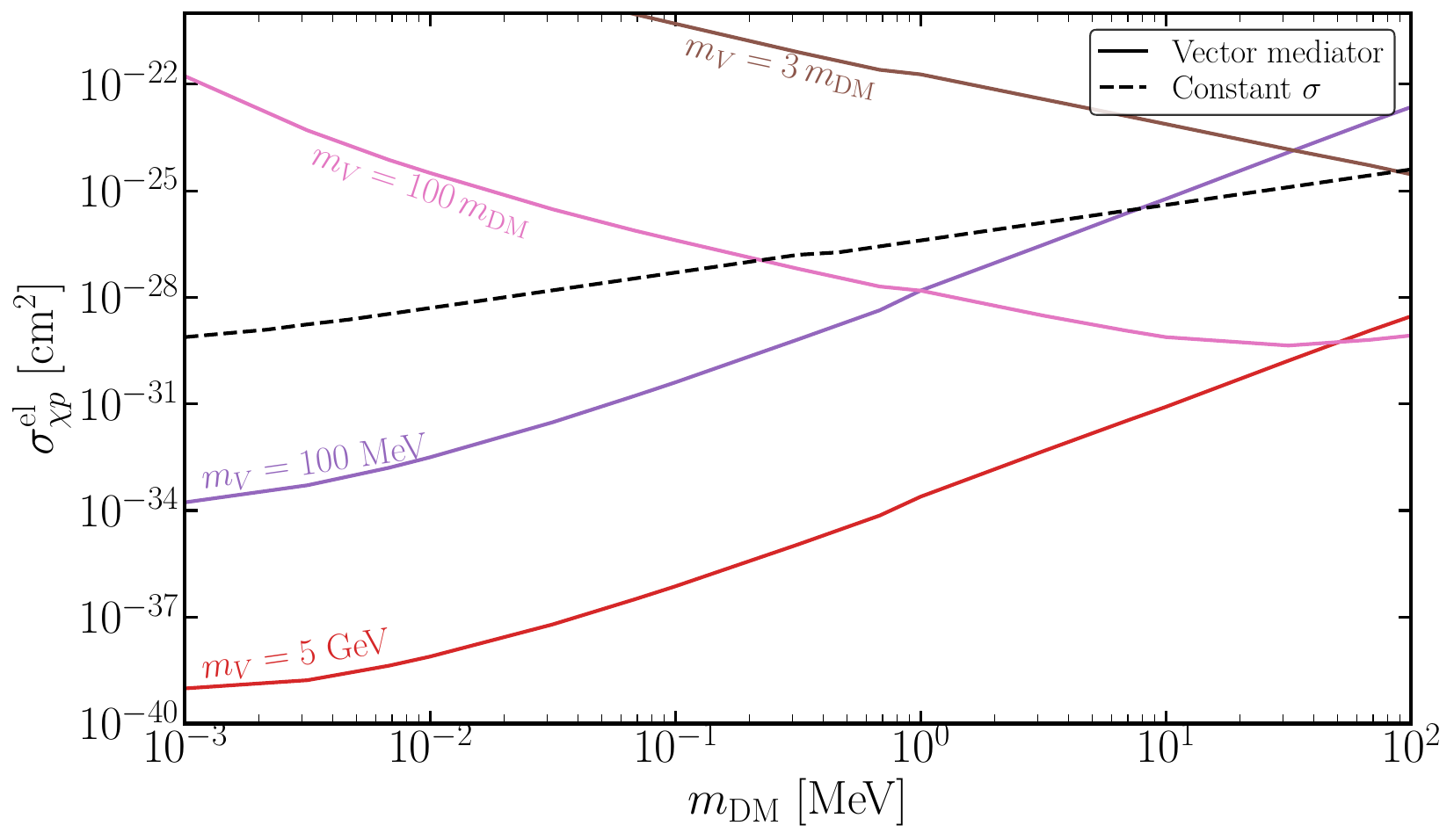}
    \includegraphics[width=0.49\textwidth]{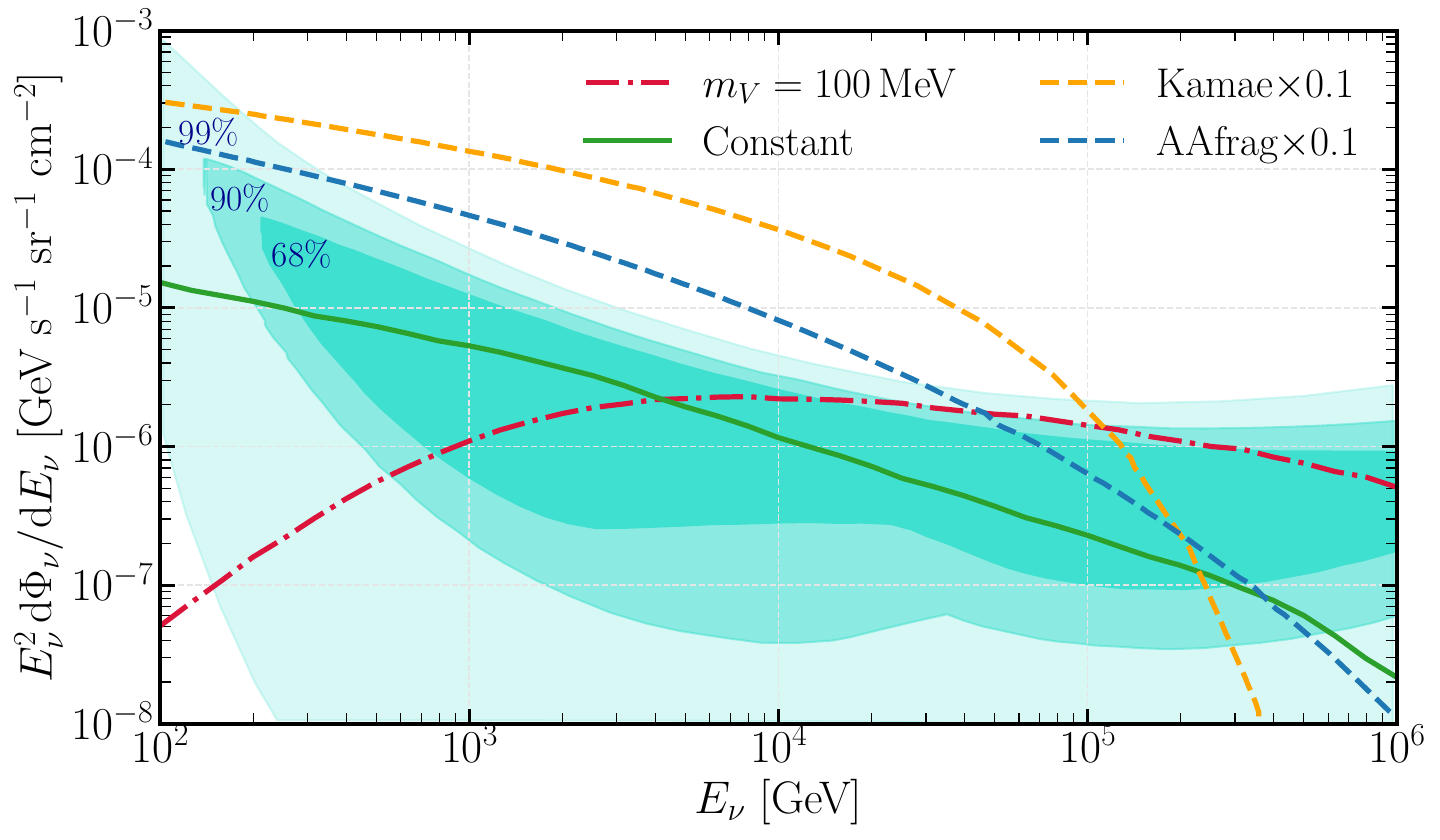}
    \caption{\emph{Left: }Derived $99\%$ C.L. upper limits on the dark matter-nucleon elastic scattering cross section as a function of the dark matter mass, utilizing the ANTARES neutrino observations in the Galactic Ridge assuming different mediator masses and a constant cross section in energy. \emph{Right: }Expected neutrino fluxes from CR-DM interactions coming from the galactic Ridge under different cross-section parameterizations for a DM mass set to 1 MeV. The constant cross section has been fixed to $\sigma_0=10^{-26}$ cm$^2$ 
    while in the case of the vector mediator the couplings are set to $G_D^2=10$. The ANTARES 68\%, 90\% and 99\% C.L. bands are shown as well~\cite{ANTARES:2022izu}.}
\label{fig:Spectra_Cross_Section} 
\end{figure*}

To explore the sensitivity of our neutrino fluxes to this choice, we also consider a fixed mass relation between dark matter and the vector mediator, and we assume the mass scalings of $m_V = 3\, m_\chi$ and $m_V = 100\, m_\chi$. In \figref{fig:Spectra_Cross_Section_Medmass} one can see how the expected neutrino fluxes change depending on the selection of the mass relation. The results are almost identical for mediator masses lower than  $m_V = 5$ GeV, while for that particular mediator mass the expected flux is reduced (even considering higher couplings) as the cross section is smaller. This is seen better in the left panel of \figref{fig:Spectra_Cross_Section}, where we show how the derived constraints change with the different mass relations. As shown there, for a fixed ratio between the DM mass and the mediator mass, the resulting cross section limits are less competitive at low DM masses and become increasingly stringent toward higher masses. In fact, they remain competitive with direct-detection constraints even at DM masses around $100$ MeV. In the very-heavy-mediator regime ($m_V \gg m_q$), which is commonly considered in DM searches, the limits become particularly strong as the elastic scattering cross section scales as $\sigma^{\rm el}_{\chi p} \propto m_V^{-4}$ (for a fixed $G_D^2$), which leads to increasingly stringent constraints as the mediator mass decreases.

Furthermore, to robustly bracket the model dependence of the resulting neutrino spectra, we compare our vector-mediator framework against two broad classes of alternative cross-section models. The first is a purely phenomenological scenario assuming an energy-independent (constant) dark matter-nucleon cross section. The second approach leverages scaled versions of standard hadronic interaction models derived from proton-proton ($p$-$p$) scattering, specifically employing the Kamae \cite{Kamae:2006bf} and AAfrag \cite{Kachelriess:2019ifk} parameterizations. Comparing the dynamically computed interactions generated by the \textsc{MadGraph5\_aMC@NLO}~\cite{Alwall:2014hca} + \textsc{Pythia 8} pipeline, described in the main text, against these empirical $p$-$p$ baseline calculations allows us to evaluate how variations in the interaction inelasticity and secondary meson multiplicities govern the final high-energy neutrino flux.

As illustrated in the right panel of \figref{fig:Spectra_Cross_Section}, these assumptions yield significantly different spectral shapes for the resulting high-energy neutrinos. For a benchmark dark matter mass of 1 MeV, we find that at lower neutrino energies ($E_\nu \sim 10^2$ GeV), the flux is dominated by the empirical $p$-$p$ models. The scaled Kamae parameterization ($\times 0.1$) yields the highest flux, followed closely by the scaled AAfrag model ($\times 0.1$). The constant cross-section assumption lies roughly a factor of 5 below Kamae, while the dynamical vector mediator model with $m_V = 100$ MeV is suppressed by 2 to 3 orders of magnitude in this low-energy regime. This highlights the importance of performing a detailed calculation of the DM-nucleon inelastic cross sections in order to avoid overestimating the predicted fluxes.

As the neutrino energy increases, the predicted fluxes from the Kamae, AAfrag, and constant cross-section models monotonically decrease. In contrast, the simulated model flux initially hardens, peaking at energies around $10^4$ GeV before turning over. At the highest observation energies, the $p$-$p$ driven spectra fall off more rapidly; the constant cross-section spectrum overtakes the Kamae and AAfrag curves around $E_\nu \sim 10^5$ GeV, and the vector mediator model eventually surpasses them around $E_\nu \sim 10^5$ GeV. Note that Kamae and AAfrag have been scaled down an order of magnitude and the coupling used in the simulations is $G_D^2=10$.

\subsection{Constant cross section}

A typical choice for the derivation of this type of limits is to parameterize the interaction with an energy-independent cross section. The cross section is then scaled to saturate the measured limit of the observable at study.
As a complementary result, we obtain cross-section limits using the same approach. We calculate the flux with the same expression as in the main text with 
\begin{equation}
    \sigma_{\rm CR\chi}\ \frac{dN_\nu}{dE_\nu} = \sigma_0\ \delta(0.05\times E_{CR}-E)
\end{equation}
and rescale $\sigma_0$ to set the limit. 

For the bounds derived using a constant inelastic cross section, we follow Ref.~\cite{Reis:2024wfy}, and map to the elastic cross section by using the relation $\sigma^{\rm el}=\frac{3}{8}\sigma^{\rm inel}$

\begin{figure*}[h!] 
    \centering 
    \includegraphics[width=0.6\textwidth]{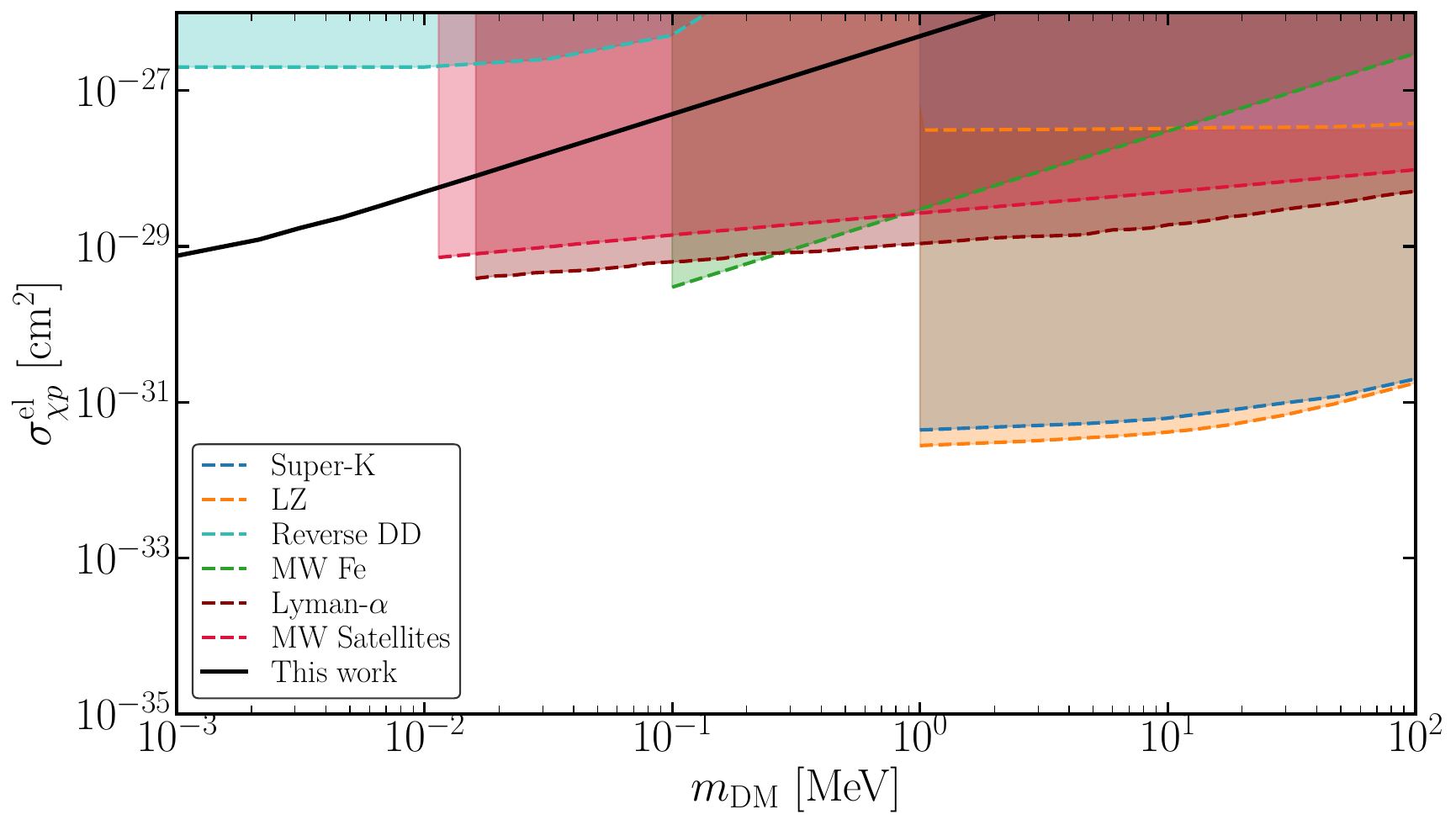}
    \caption{Derived $99\%$ C.L. upper limits on the dark matter-nucleon elastic scattering cross section as a function of the dark matter mass, utilizing the ANTARES neutrino observations in the Galactic Ridge and assuming a constant cross section in energy. In addition we show limits coming from Reverse Direct Detection~\cite{Cappiello:2018hsu}, Lyman-$\alpha$ forest observations~\cite{Rogers:2021byl}, Milky Way satellite galaxy population~\cite{DES:2020fxi}. We also show the limits coming from CRDM obtained with LZ~\cite{Maity:2022exk} and Super-Kamiokande~\cite{Super-Kamiokande:2022ncz}, from cosmic ray iron spallation in the Milky Way~\cite{Meighen-Berger:2025hrq}.
    Other constraints that might be of interest are discussed in the main text. }
\label{fig:Constant_Bounds} 
\end{figure*}

In Fig.~\ref{fig:Constant_Bounds} we show our results and compare with other constraints, including CRDM limits from LZ~\cite{Maity:2022exk} and Super-Kamiokande~\cite{Super-Kamiokande:2022ncz}, and bounds based on cosmic ray spallation~\cite{Lu:2023aar, Meighen-Berger:2025hrq} and cosmic ray spectra modification in the Milky Way~\cite{Cappiello:2018hsu}. Cosmological bounds from the Milky Way satellite population~\cite{DES:2020fxi} and the Lyman-$\alpha$ forest~\cite{Rogers:2021byl} apply here in much the same way that they do for a heavy vector mediator, and are shown alongside the above mentioned constraints based on cosmic ray interactions. If one considers the existence of spikes both in the dark matter profile of the Milky Way or in NGC1068, stronger constraints apply, see \cite{Meighen-Berger:2025hrq, Herrera:2023nww,Gustafson:2024aom,Gustafson:2025dff}.

\section{Astrophysical uncertainties}

\subsection{Cosmic-ray distribution}

The spatial and spectral distribution of cosmic rays (CRs) in the inner Galaxy represents a primary source of astrophysical uncertainty when estimating the secondary neutrino flux from CR-DM interactions. To quantify this systematic effect, we compare predictions derived from different CR transport scenarios.

The most conservative baseline is the Base-Minimal scenario. This model assumes a spatially uniform diffusion coefficient across the entire Galactic disk and halo, with parameters tuned strictly to reproduce local CR measurements at Earth. While this setup provides a solid anchor locally, it generally struggles to reproduce the morphology and spectral hardening of the diffuse $\gamma$-ray emission observed towards the Galactic Center.

To address these shortcomings, we heavily rely on the $\gamma$-optimized transport models~\cite{DelaTorreLuque:2022ats}. These models introduce a spatially dependent, non-uniform diffusion coefficient that decreases towards the inner Galaxy. This physically motivated modification naturally produces a harder CR spectrum and a higher CR density in the Galactic Ridge, allowing it to simultaneously fit high-precision local CR data and the diffuse $\gamma$-ray measurements across a vast range of energies \cite{DeLaTorreLuque:2025zsv}. The ``Min'' and ``Max'' setups serve to bracket the uncertainties within these non-uniform models, largely stemming from slightly different fits to local CR injection spectra.

\begin{figure*}[h!] 
    \centering 
    \includegraphics[width=0.6\textwidth]{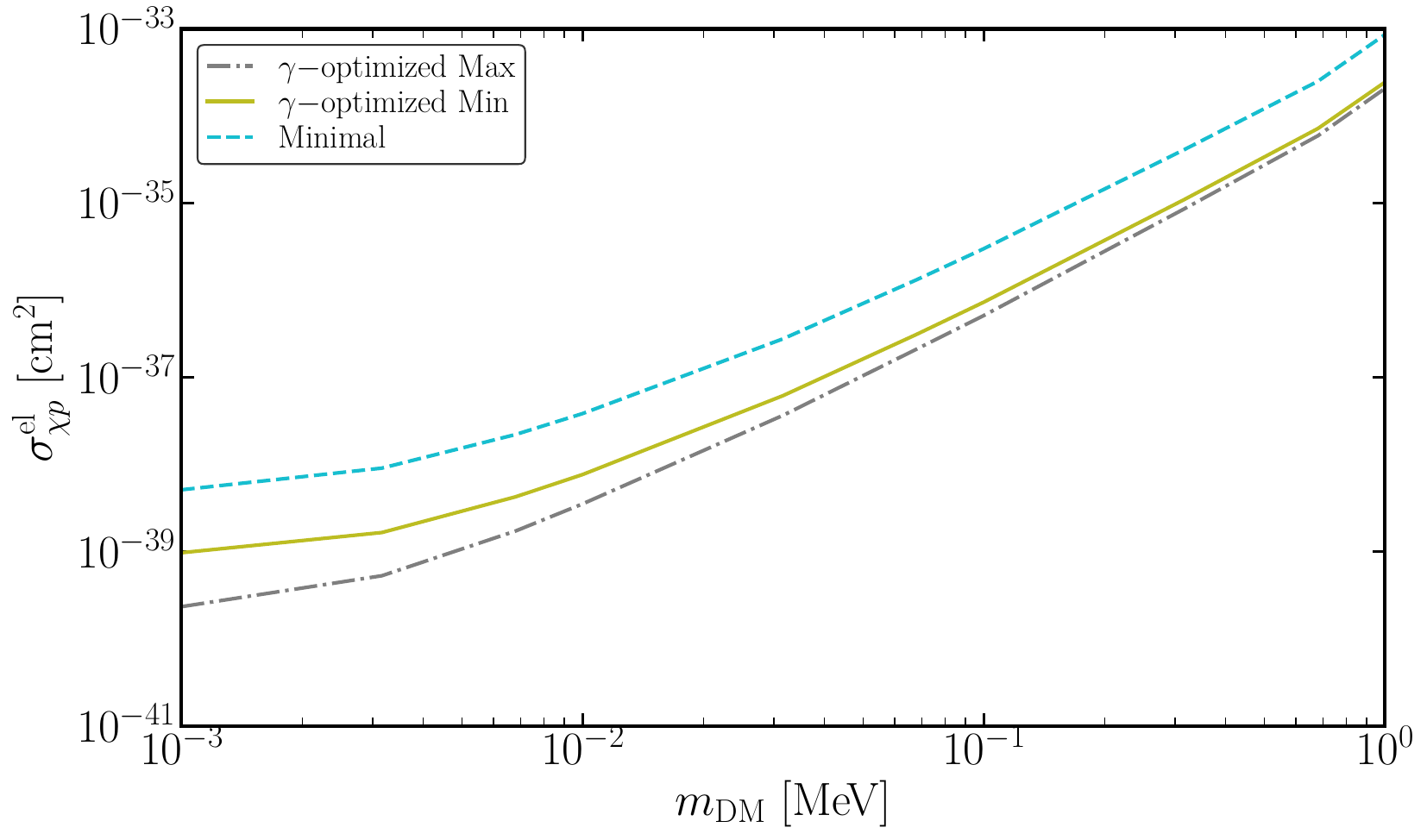}
    \caption{Derived $99\%$ C.L. upper limits on the dark matter-nucleon elastic scattering cross section as a function of the dark matter mass, utilizing the ANTARES neutrino observations in the Galactic Ridge assuming a $m_V=5\ \rm{GeV}$ mediator for different galactic Cosmic Ray models.}
\label{fig:CR_Uncs} 
\end{figure*}

The choice between these propagation models has a profound impact on the derived dark matter constraints. The quantitative scaling of this systematic uncertainty is depicted in \figref{fig:CR_Uncs}, assuming our Dirac DM fermion benchmark model with $m_V = 5\ \rm  GeV$. The two $\gamma$-optimized setups yield very comparable constraints across the entire dark matter mass range considered, effectively bracketing the underlying systematic variation related to the inner Galaxy density: the ``Max'' parameterization provides marginally stronger constraints for $m_\chi \lesssim 0.1$ MeV, while for larger masses the results almost overlap.

Conversely, the Base-Minimal scenario yields noticeably softer bounds. Because the Minimal diffusion model predicts a less dense, softer population of high-energy protons in the Galactic Ridge, it leads to a significantly lower predicted neutrino yield for a given DM-nucleon cross section. For $m_\chi \lesssim 0.1$ MeV, the resulting cross-section constraints are approximately an order of magnitude weaker than those derived from the $\gamma$-optimized setups. At higher masses, its bounds draw closer to the $\gamma$-optimized ones, but remain consistently less competitive across the parameter space.

\subsection{Dark matter density profiles}

The predicted neutrino flux from interactions between cosmic rays and dark matter is contingent upon the spatial distribution of dark matter within the Galactic halo. As our fiducial model, we adopt the Navarro-Frenk-White (NFW) profile~\cite{Navarro:1995iw}, which features a cusp in the inner Galaxy. To properly assess the robustness of the derived limits and account for astrophysical uncertainties, we explore alternative dark matter density profiles, comparing our baseline results with those obtained assuming Burkert~\cite{Burkert:1995yz}, Einasto~\cite{Einasto:1965czb} and Moore~\cite{Moore:1999nt} profiles.

\begin{figure*}[h!] 
    \centering 
    \includegraphics[width=0.6\textwidth]{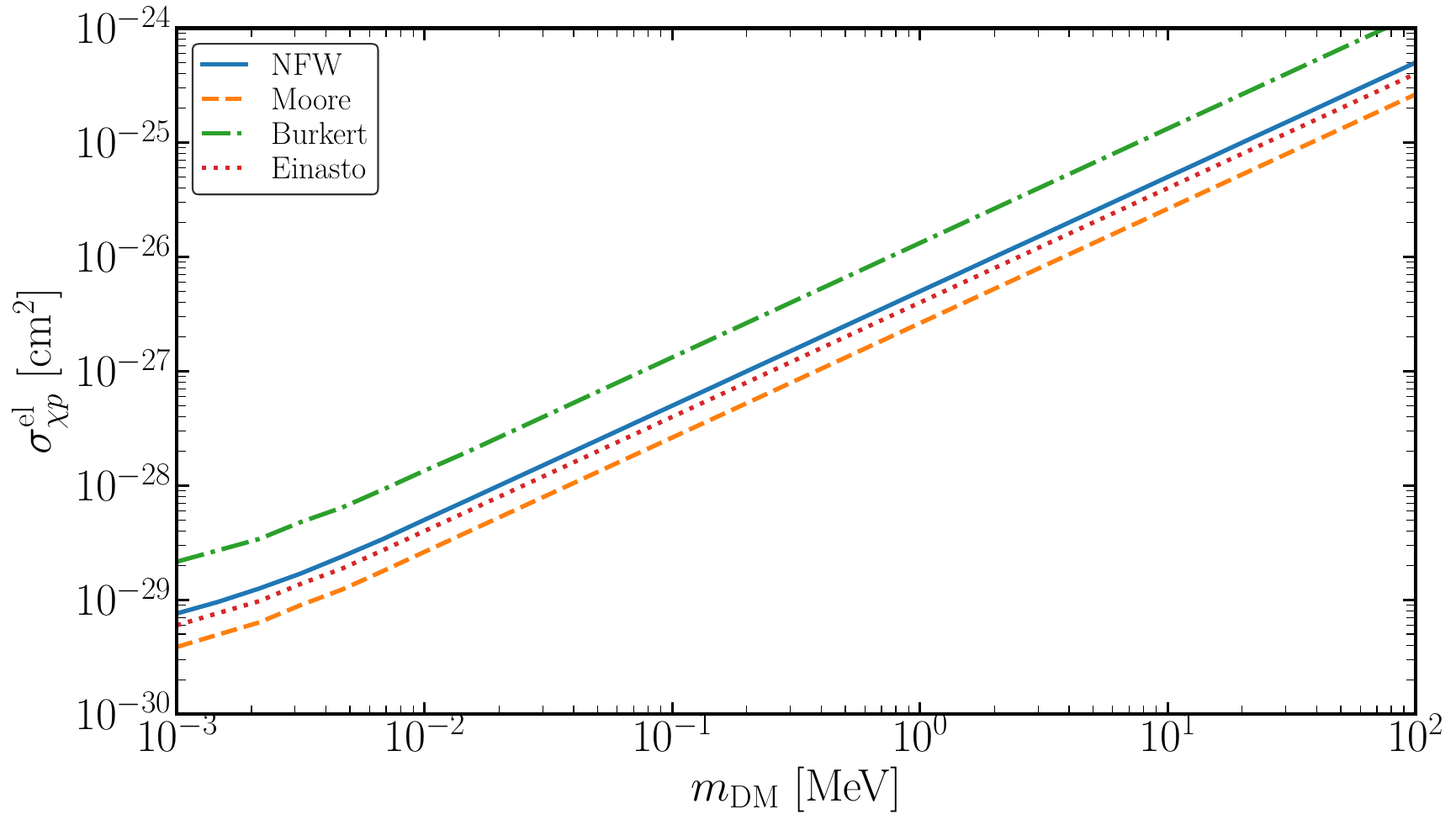}
    \caption{Impact of the dark matter density profile uncertainties on the derived cross-section upper limits, assuming a constant interaction cross section.}
\label{fig:DMProf_Uncs} 
\end{figure*}

Because the neutrino flux produced in deep inelastic scattering of proton cosmic rays with dark matter scales linearly with the dark matter density, the choice of the profile has a direct impact on the resulting cross-section bounds. As illustrated in \figref{fig:DMProf_Uncs}, which displays the derived upper limits under the assumption of a constant dark matter-proton cross section, the resulting constraints scale uniformly with dark matter mass across all profile models. Profiles with steeper inner slopes, such as the Moore profile, provide the strongest constraints, as the assumed dark matter density increases rapidly toward the Galactic center. In contrast, adopting a cored profile like Burkert generates a much more conservative dark matter density in the inner Galaxy, resulting in constraints roughly an order of magnitude weaker.

Between these two extremes lie the Einasto and NFW profiles, which are both theoretically well-motivated standard parameterizations. These two models yield very similar constraints in the central parsecs of the Galaxy, with the limits from the NFW profile lying marginally weaker than those derived from the Einasto profile. Overall, our baseline choice of the NFW profile in the main text represents a robust and moderate middle ground for assessing empirical sensitivity without overcommitting to the steep cusp of the Moore profile.

\section{Particle physics uncertainties}

\subsection{PDF systematic uncertainty}

The deep inelastic cross section $\sigma_{\rm inel}(E_p)$ is weighted by the proton parton distribution functions (PDFs), which are extracted from global QCD analyses and carry both experimental and methodological uncertainties~\cite{Lai:2010vv,NNPDF:2012nn,NNPDF:2013qed,Cruz-Martinez:2024nnpdf}.  At the highest cosmic-ray energies our analysis probes Bjorken-$x$ values as low as $x\sim 10^{-8}$, a regime in which PDFs are poorly constrained by present data.  The choice of PDF set can therefore be a non-negligible source of systematic error.

To quantify this uncertainty we have recomputed the proton--DM cross section and the associated neutrino yield for a representative set of four modern PDF determinations: the LO set \texttt{NNPDF23\_lo\_as\_0130\_qed} with QED corrections~\cite{NNPDF:2013qed}, which we adopt as our benchmark; the NLO set \texttt{CT10nlo} without QED corrections~\cite{Lai:2010vv}; the NLO set \texttt{NNPDF23\_nlo\_as\_0119\_qed} with QED corrections~\cite{NNPDF:2013qed}; and the modern LO set \texttt{NNPDF40\_lo\_as\_01180} optimized for Monte Carlo applications~\cite{Cruz-Martinez:2024nnpdf}.  The comparison spans four incident proton energies that bracket the range dominating the line-of-sight integral for the Galactic Ridge neutrino flux.

\textit{Neutrino yield.}  For all four PDF sets the differential neutrino energy distribution $dN_\nu/dE_\nu$ at fixed $E_p$ agrees with the benchmark template to within the statistical precision of the \textsc{MadGraph5\_aMC@NLO}+\textsc{Pythia~8} simulation chain.  This stability is expected: the shape of the neutrino spectrum is governed by the hadronization and decay chain, which in turn depend on the momentum transfer $Q^2$ and the Bjorken-$x$ of the initiating hard scatter.  When the same generator setup (parton shower, hadronization model, decay tables) is employed, the relative parton kinematics are insensitive to the overall PDF normalization.  The template $dN_\nu/dE_\nu(E_p)$ can therefore be regarded as PDF-independent for the purpose of the present analysis.

\begin{figure*}[h!] 
    \centering 
    \includegraphics[width=0.6\textwidth]{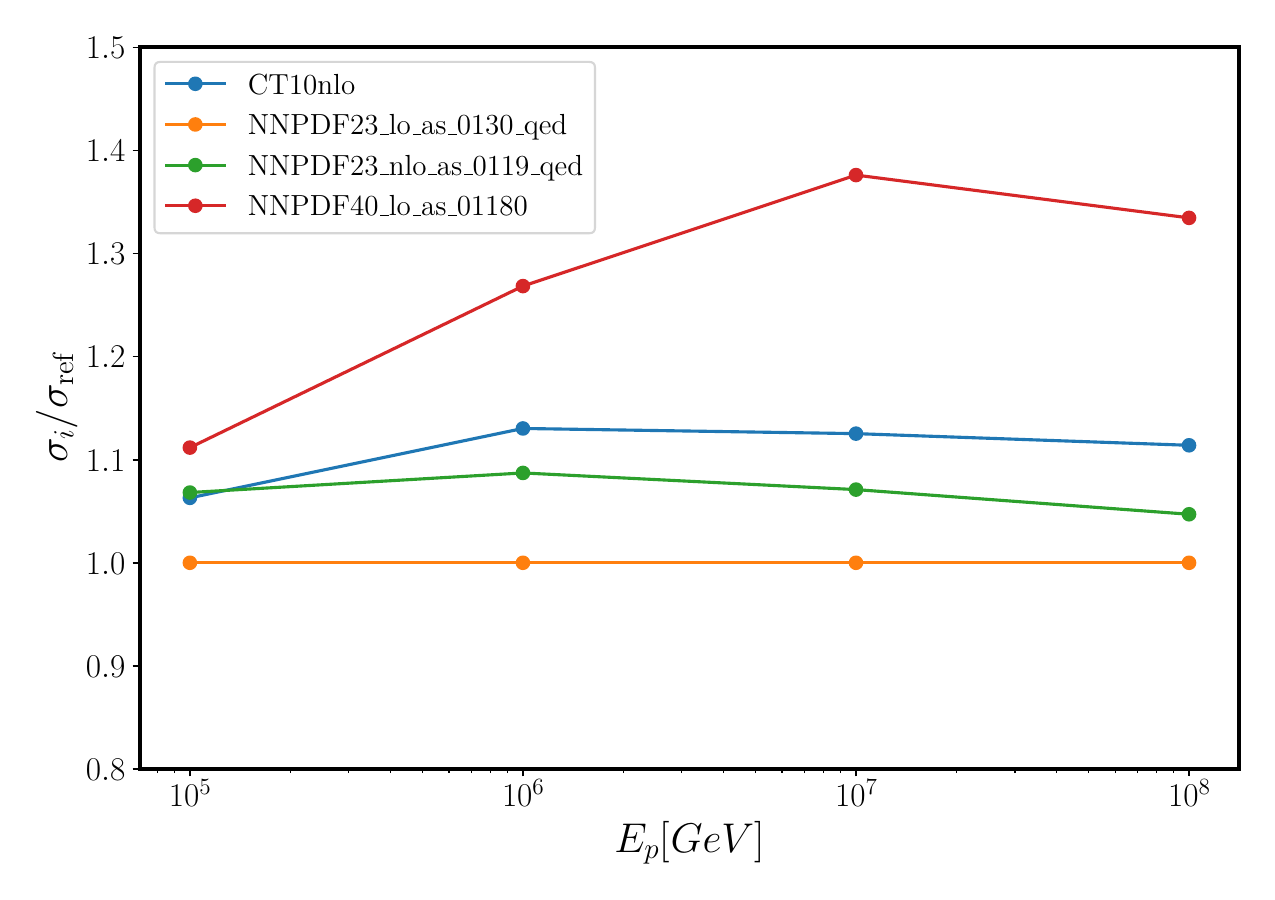}
    \caption{Ratio of the total cross section calculated with \textsc{MadGraph5\_aMC@NLO} for different proton PDF over the one calculated with the benchmark model.}
\label{fig:PDF_ratio} 
\end{figure*}

\textit{Total inelastic cross section.}  In contrast, the total cross section $\sigma_{\rm inel}(E_p)$ is directly proportional to the parton luminosity and therefore reflects the differences among the PDF sets.  Fig.~\ref{fig:PDF_ratio} displays the ratio $\sigma_{\rm PDF}/\sigma_{\rm benchmark}$ for each set as a function of the incident proton energy $E_p$.  The NLO set \texttt{NNPDF23\_nlo\_as\_0119\_qed} is systematically larger than the benchmark by roughly $5\%$ across the full energy range, while \texttt{CT10nlo} exceeds it by roughly $10\%$ with an almost flat energy dependence.  The modern LO set \texttt{NNPDF40\_lo\_as\_01180} is always larger than the benchmark, and the excess grows from approximately $10\%$ at $E_p = 10^5$~GeV to roughly $30\%$ at $10^8$~GeV.  All tested PDF sets therefore yield cross sections that are larger than or equal to our benchmark choice, so that the benchmark \texttt{NNPDF23\_lo\_as\_0130\_qed} provides the most conservative signal prediction among the sets examined.

The energy-dependent growth of the \texttt{NNPDF40\_lo\_as\_01180} ratio can be traced to the fact that the NNPDF4.0 family incorporates more recent small-$x$ data from HERA and the LHC, which drives a modest hardening of the gluon and sea-quark distributions at very low $x$~\cite{Cruz-Martinez:2024nnpdf}.  At the extreme centre-of-mass energies probed by high-energy cosmic rays, this enhanced small-$x$ gluon propagates into a correspondingly larger DIS cross section.

\section{Prospects for next generation of neutrino instruments}
\label{sec:neutrino_prospects}

In addition to the ANTARES Galactic Ridge limits discussed in the main text, we derive constraints using the current IceCube observation of the diffuse Galactic Plane neutrino flux and project the sensitivity reach of the upcoming IceCube-Gen2 and KM3NeT detectors. The resulting limits are shown in Fig.~2 of the main text; here we detail the methodology.

\subsection{IceCube Galactic Plane}

The IceCube Collaboration has reported a diffuse neutrino flux from the Galactic Plane ($|l| \leq 180^\circ$, $|b| \leq 90^\circ$). Unlike the model-agnostic ANTARES Ridge measurement, the IceCube result is more sensitive but template-dependent. Among the available templates ($\pi^0$, KRA$_\gamma^5$, KRA$_\gamma^{50}$), we adopt the $\pi^0$ model as it yields the most conservative flux, ensuring our derived limits are robust.

Our predicted DM-induced neutrino flux is computed over the Galactic Ridge region ($|l| < 30^\circ$, $|b| < 2^\circ$), while the IceCube reported fit covers the full sky. To bridge this mismatch, we rescale the Ridge DM flux to an equivalent all-sky flux using a template-dependent factor derived from the $\pi^0$ model itself:
\begin{equation}
    \mathcal{R}_{\pi^0} \equiv 
    \frac{\int_{\rm all\;sky} d\Omega\,\phi_{\pi^0}(l,b,E_\nu)}
         {\int_{\rm Ridge}   d\Omega\,\phi_{\pi^0}(l,b,E_\nu)} 
    \approx 21.5\,,
    \label{eq:R_pi0}
\end{equation}
so that the effective all-sky DM flux is obtained as
$\phi_{\rm DM}^{\rm all\;sky} = (4\pi/\mathcal{R}_{\pi^0})\,\phi_{\rm DM}^{\rm Ridge}$.
This rescaling is considerably more conservative than a purely geometric solid-angle ratio ($4\pi/\Delta\Omega_{\rm Ridge}\approx 172$), because the $\pi^0$ template is strongly concentrated along the Galactic plane---precisely where the Ridge is located---meaning the Ridge already captures a substantial fraction of the total expected Galactic signal.

For each DM mass, the energy-dependent all-sky flux $\phi_{\rm DM}(E_\nu)$ is compared to the IceCube $\pi^0$ best-fit flux (converted from per-flavor to all-flavor by multiplying by 3) in the energy range $1$--$70$~TeV, using the IceCube energy bin edges. The $G_D^2$ coupling limit is set by the most constraining bin:
\begin{equation}
    G_D^2 \leq G_{D,\rm ref}^2 \times 
    \min_i\left[
        \frac{\int_{E_i}^{E_{i+1}} \phi_{\rm IC}(E_\nu)\,dE_\nu}
             {\int_{E_i}^{E_{i+1}} \phi_{\rm DM}(E_\nu)\,dE_\nu}
    \right]\,,
    \label{eq:bin_limit}
\end{equation}
with $G^2_{D,\rm ref}=1$. The bin integrals are evaluated by trapezoidal integration. The resulting bound on $G_D^2$ is converted to an elastic DM--proton cross section via Eq.~(4) of the main text.

\subsection{IceCube-Gen2}

IceCube-Gen2, the planned high-energy extension of IceCube, will feature an instrumented volume approximately five times larger, yielding a corresponding increase in the effective area for muon neutrinos~\cite{IceCube-Gen2:2020qha}. For a diffuse flux measurement in the background-dominated regime---as is the case for the Galactic Plane neutrino search---both the signal and the atmospheric-neutrino background scale linearly with the effective area, so the statistical precision on the extracted flux normalization improves as $1/\sqrt{A_{\rm eff}}$.  We therefore conservatively estimate the Gen2 reach by rescaling the current IceCube best-fit flux by a factor $\sqrt{5}\approx 2.24$:
\begin{equation}
    \phi_{\rm Gen2}(E_\nu) = \phi_{\rm IC}(E_\nu)\,/\,2.24\,,
    \label{eq:gen2_scaling}
\end{equation}
and applying the same bin-by-bin limit-setting procedure of Eq.~\eqref{eq:bin_limit}.  This assumes the same spectral templates and energy range as current IceCube; in practice, the broader energy reach of Gen2 (extending to PeV energies) and its improved angular resolution will further enhance the sensitivity, particularly for DM masses $m_\chi\lesssim 10$~keV whose neutrino spectra peak at higher energies.  The projected Gen2 limits are therefore conservative.

\subsection{KM3NeT}

KM3NeT, currently under construction in the Mediterranean Sea but already operational and taking data, comprises the ARCA detector optimized for high-energy ($\gtrsim$~TeV) neutrino astronomy~\cite{LeBreton:2019lpq}.  ARCA's effective area for muon neutrinos is expected to be $10$--$100$ times larger than that of ANTARES~\cite{Bozza:2023vO}.  As argued above, the sensitivity to a diffuse flux improves with the square root of the effective area in the background-limited regime.  Taking the lower bound of the anticipated range ($A_{\rm eff}^{\rm ARCA}\approx 10\,A_{\rm eff}^{\rm ANTARES}$) yields a conservative sensitivity gain of $\sqrt{10}\approx 3.2$, which we round to $3.5$:
\begin{equation}
    \Phi_{{\rm KM3NeT},\,i}^{99\%} = \Phi_{{\rm ANTARES},\,i}^{99\%}\,/\,3.5\,.
    \label{eq:km3net_scaling}
\end{equation}
The ANTARES 99\%~C.L.\ upper limits per energy bin~\cite{ANTARES:2022izu} are directly rescaled, and the same limit-setting procedure is applied.  We stress that this scaling is deliberately conservative; the realistic improvement will be larger owing to KM3NeT's superior energy and directional resolution, as well as the possibility of a combined ARCA+ORCA analysis.

\end{document}